\documentclass[conference]{IEEEtran}
\IEEEoverridecommandlockouts
\usepackage{algorithm}
\usepackage{algorithmic}
\usepackage{amsmath}
\usepackage{array}
\usepackage{multirow}
\usepackage[capitalize]{cleveref}
\usepackage{balance}
\usepackage{subcaption}
\usepackage{censor}
\usepackage{float}
\usepackage{hhline}
\usepackage{dblfloatfix}
\usepackage{xcolor}
\usepackage{graphicx}
\usepackage{fancyhdr}
\usepackage{tikz}

\def\BibTeX{{\rm B\kern-.05em{\sc i\kern-.025em b}\kern-.08em
    T\kern-.1667em\lower.7ex\hbox{E}\kern-.125emX}}

\begin{document}

\title{A Hierarchical Optimization Framework Using Deep Reinforcement Learning for Task-Driven Bandwidth Allocation in 5G Teleoperation}

\author{\IEEEauthorblockN{
\large Narges Golmohammadi, Madan Mohan Rayguru and Sabur Baidya\\
\normalsize Department of Computer Science and Engineering, University of Louisville, KY, USA}
\normalsize {e-mail:  narges.golmohammadi@louisville.edu, madanmohan.rayguru@louisville.edu, sabur.baidya@louisville.edu}
\vspace{-4mm}
}


\iffalse % commenting for double-blind submission

\author{Narges Golmohammadi}
\affiliation{%
  \institution{University of Louisville}
  \city{Louisville}
  \state{Kentucky}
  \country{USA}
\email{narges.golmohammadi@louisville.edu}
  }

\author{Madan Mohan Rayguru}
\affiliation{%
  \institution{University of Louisville}
  \city{Louisville}
  \state{Kentucky}
  \country{USA}
\email{madanmohan.rayguru@louisville.edu}
  }
  
  \author{Sabur Baidya}
\affiliation{%
  \institution{University of Louisville}
  \city{Louisville}
  \state{Kentucky}
  \country{USA}
\email{sabur.baidya@louisville.edu}
  }

  \fi

% IEEE copyright notice - ONLY FOR ARXIV UPLOAD

\maketitle
\pagestyle{fancy}
\thispagestyle{fancy}
\renewcommand{\headrulewidth}{0pt}  % Remove header line

\fancyhf{}

% HEADER (updated per your request)
\fancyhead[C]{
    \small This article has been accepted for publication in the INFOCOM Workshop (NetRobiCS 2025)
}

% FOOTER (unchanged)
\fancyfoot[C]{
    \begin{tikzpicture}[remember picture, overlay]
        \node[anchor=south,yshift=10pt] at (current page.south) {
            \parbox{\textwidth}{
                \centering
                \footnotesize
                \textcopyright~2025 IEEE. Personal use of this material is permitted. Permission from IEEE must be obtained for all other uses, in any current or future media, including reprinting/republishing this material for advertising or promotional purposes, creating new collective works, for resale or redistribution to servers or lists, or reuse of any copyrighted component of this work in other works.
            }
        };
    \end{tikzpicture}
}

\begin{abstract}
\textcolor{black}{The evolution of 5G wireless technology has revolutionized connectivity, enabling a diverse range of applications. Among these are critical use cases such as real-time teleoperation, which demands ultra-reliable low-latency communications (URLLC) to ensure precise and uninterrupted control, and enhanced mobile broadband (eMBB) services, which cater to data-intensive applications requiring high throughput and bandwidth. In our scenario, there are two queues—one for eMBB users and one for URLLC users. In teleoperation tasks, control commands are received in the URLLC queue, where communication delays occur. The dynamic index (DI) controls the service rate, affecting the telerobotic (URLLC) queue. A separate queue models eMBB data traffic. Both queues are managed through network slicing and application delay constraints, leading to a unified Lagrangian-based Lyapunov optimization for efficient resource allocation. We propose a DRL-based hierarchical optimization framework that consists of two levels. At the first level, network optimization dynamically allocates resources for eMBB and URLLC users using a Lagrangian functional and an actor-critic network to balance competing objectives. At the second level, control optimization fine-tunes the best gains for robots, ensuring stability and responsiveness in network conditions. This hierarchical approach enhances both communication and control processes, ensuring efficient resource utilization and optimized performance across the network.}

\end{abstract}

\begin{IEEEkeywords}
%\keywords{
Tactile Internet, Telerobotics, 5G, Network Slicing, Deep Reinforcement Learning, Hierarchical Optimization %}
\end{IEEEkeywords}



\vspace{-2mm}
\section{Introduction}
\vspace{-1mm}
The rapid advancement in sensing, processing and control technologies have enabled exponential growth in robotic applications~\cite{leng2022industry, demir2019industry, khalid2016methodology} in recent times.
%This is also corroborated by the demands of new and emerging applications, e.g., Industry 5.0 ~\cite{leng2022industry, demir2019industry}, autonomous vehicles~\cite{martinez2018autonomous}, healthcare and medical robots~\cite{taylor2016medical, riek2017healthcare}, and other critical cyber-physical robotic applications~\cite{khalid2016methodology}. 
Although autonomous operations of robots are prevalent in many mission-critical applications, safety-critical applications involving robots often need teleoperation with human supervision. For example, robots used in medical surgery are usually teleoperated under the supervision of a trained physician~\cite{choi2018telesurgery}. As typically the users and/or the robots are mobile in these applications, the teleoperation is enabled by 5G wireless communications. 

 \textcolor{black}{In the context of 5G, telerobotic applications are categorized as Tactile Internet applications, which encompass both Ultra-Reliable Low-Latency Communications (URLLC) and enhanced Mobile Broadband (eMBB)~\cite{khan2022urllc, ali2021urllc}. These applications require stringent latency and reliability standards, with URLLC ensuring low-latency control commands from operators to robots, while eMBB is responsible for transmitting the latest system status and video feedback, enabling operators to provide accurate and effective commands.
%Telerobotic users, therefore, utilize both URLLC and eMBB slices to support their unique needs. Additionally, other eMBB users, such as those requiring high-definition video streaming or data-intensive services, share the same 5G wireless resources. To maintain the stability and reliability of the system, intelligent and adaptive bandwidth allocation is essential.
Given their distinct Quality of Service (QoS) requirements, 5G networks employ network slicing. This allows for dedicated slices—one optimized for URLLC to handle control commands and another slice designed for eMBB to manage video feedback and system status updates—ensuring optimal performance for both applications.
The dynamic nature of wireless environments and the varying demands of telerobotic operations introduce challenges in maintaining stable performance, control, and overall system reliability. Through network slicing, 5G networks can effectively address these diverse needs, guaranteeing robust and efficient service for all users.}

\textcolor{black}{This work aims to address a hierarchical 2-level optimization, where the upper-level joint Lyapunov optimization ensures performance objectives at both the network and application levels while maintaining queue stability. We consider two queues -- one for enhanced Mobile Broadband (eMBB) users and another for Ultra-Reliable Low-Latency Communications (URLLC) users. Both queues have arrival rates modeled by Poisson processes.
For URLLC users, the departure rate is determined by the Dexterity Index (DI) combined with throughput, ensuring low-latency, reliable communication. For eMBB users, the departure rate is governed by the overall system throughput.
%, supporting high data rates and efficient resource allocation.
At the lower level, a robust closed-loop control is maintained using Razumikhin-based linear quadratic regulator (LQR) \cite{rayguru2020} to ensure stable control gains that meet delay constraints in dynamic environments. Given the stochastic nature of the upper-level optimization, occasional violations of delay constraints may occur, particularly when reliability is low. To handle these cases, we adapt nominal control gains to ensure stable performance.
Through numerical simulations, we demonstrate the effectiveness of our framework, which includes efficient resource allocation through slicing, stable network performance, and robust control gains for telerobotic users, meeting the specified objectives.
}

In summary, our contributions are as follows:

\begin{itemize}\itemsep 2pt
\vspace{-1mm}
\item \textcolor{black}{A hierarchical optimization framework is developed that employs a deep reinforcement learning model to allocate resources across slices, and a Razumikhin-based LQR controller to derive robust optimal control gains based on network delay.} 
%The DRL layer addresses the joint Lyapunov optimization problem by balancing objectives such as QoS, constraint satisfaction, and queue stability, while adapting to dynamic network conditions and user demands. Meanwhile, the lower-level LQR control leverages network delay as a parameter for closed-loop gain computation, enhancing robot control precision and task execution in the face of fluctuating network performance. 
%This layered structure allows for optimized resource allocation and control at both the network and application levels, ensuring that QoS requirements for teleoperation are met under varying conditions.
    \item %We formulate a joint optimization of Network level QoS and application level control for tactile Internet applications with coexisting URLLC and eMMBB users. Our novel formulation handles satisfying the low level control dexterity while  simultaneously aiming to stabilize the queue due to varying user and network dynamics.

\textcolor{black}{We present a Lyapunov-based joint optimization framework tailored for tactile Internet applications, which uses a dual-queue model to address the distinct demands of URLLC-driven telerobotic control and eMBB data traffic. It leverages 5G network slicing and subcarrier resource allocation in the radio access network (RAN)
%to provide isolation between the two service types while 
for adapting to fluctuations in network conditions and user demands.
%By integrating a conventional queue for eMBB traffic and a specialized queue for telerobotic users—whose dynamics are influenced by user behavior and the robot's task execution efficiency—the framework ensures simultaneous stability, QoS for telerobotics, and efficient eMBB service delivery.
}

    \item %To solve the problem efficiently, we develop a DRL based framework with actor-critic network and formulate a reward function that handles balancing the performance objective, constraints and queue stability during the training. With experiments in varying condition, we show the efficient and robust decisions made by the DRL framework in real-time. 
   \textcolor{black}{Unlike standard DRL methods, which are typically designed for unconstrained Lyapunov optimization, we integrate a Lagrangian function to account for network constraints directly within the optimization process. %The actor-critic network is designed with a 
   The reward function we design effectively balances multiple competing goals, including quality of service, adherence to constraints, and queue stability. 
   %Extensive simulations across varied real-world scenarios demonstrate the robustness and effectiveness of our DRL framework in optimizing data rate decisions in real-time.
   }

\end{itemize}

%\begin{figure*}[!t]
%\centerline{\includegraphics[width=0.65\linewidth]{images/sliced_telerobotics.jpg}}
%\vspace{2mm}
%\caption{\centering {5G Network Slicing with optimized resource allocation supporting Telerobotic Applications\\ coexisting with eMBB Applications}}
%\label{fig1:topology}
%\vspace{-2mm}
%\end{figure*}

%%%%%%%%%%%%%%%%%%%%
\section{Related Work}

%\vspace{-1mm}

%Increasingly reliable low-latency communications (URLLC) and enhanced mobile broadband (eMBB) are now possible with 5G technology. For a unified network to accommodate both of these services, sophisticated resource allocation methods need to be employed that can adapt to varying Quality-of-Service (QoS) demands while maintaining system stability.

Dynamic resource allocation is crucial to balance competing needs of eMBB and URLLC in 5G networks. \cite{10.23919/icn.2022.0011} and \cite{10.1109/camad.2019.8858433}  highlight techniques that optimize resource use to achieve low latency and high reliability for URLLC and high bandwidth for eMBB. Similarly, Korrai et al. %\cite{10.1109/camad.2019.8858433} emphasize network slicing to multiplex these services for concurrent operation without degrading performance.
Chen et al. \cite{10.1109/mcom.2018.1701178} underscore the critical role of ultra-low latency in supporting mission-critical applications.
For diverse 5G service demands, network slicing is essential for isolating and guaranteeing performance. Popovski et al. \cite{10.1109/access.2018.2872781} focused to ensure URLLC’s stringent latency targets do not impact eMBB’s high data needs. Santos et al. \cite{10.1109/lcomm.2019.2959335} explore max-matching diversity channel allocation within network slicing, demonstrating effective load balancing. %for both services.

In addition to resource allocation, queuing models for URLLC and eMBB traffic help maintain QoS. Makeeva \cite{10.3390/math11183925} examines the impact of URLLC on eMBB quality, highlighting the delicate balance needed in shared networks, while Zhang et al. \cite{10.1109/lwc.2020.3046628} propose a stochastic optimization framework to jointly schedule both traffic types effectively.
 In \cite{kasgari2018stochastic, samanipour2023stability}, the authors investigate power allocation to different 5G slices using Lyapunov optimization, but do not consider the capacity constraint for URLLC users and the control dynamics. In~\cite{golmohammadi2024lyapunov}, the authors formulate a Lyapunov optimization of eMBB and URLLC slices to minimize the control drift error. %However, this work does not consider the control performance and the queue dynamics of both types of users and their proposed solution is not suitable for real-time implementation.% in the practical 5G network.
 
Deep reinforcement learning (DRL) shows promise in optimizing resource allocation under changing conditions. Filali et al. \cite{10.48550/arxiv.2202.06435} showcase DRL’s potential to adapt resource allocations dynamically, enhancing performance for both URLLC and eMBB, a sentiment echoed by Zhang et al. \cite{10.1109/access.2019.2917751} in machine learning-based scheduling approaches.

However, these works do not consider the control performance and the network queue dynamics of both types of users and their proposed solution is not suitable for real-time implementation.
%The coexistence of URLLC and eMBB in 5G networks requires advanced resource allocation techniques, network slicing, and intelligent scheduling. 
Our proposed research lays a robust foundation for understanding and addressing these challenges in dynamic environments, with advanced resource allocation techniques, network slicing, and intelligent scheduling.
%guiding future advancements in telecommunications.

%\section{Preliminaries: Autonomous robotics vs Telerobotics}

\section{System Model}
%\textcolor{red}{
%- It should mention the network model, application model (eMBB and URLLC), and control model with dexterity index as well
%- Madan: Please add text here, especially the control model. Narges added some equations; you can update the equations and add text\\}

\subsection{Application Scenario}

\vspace{-1mm}
We consider a scenario of a 5G network which is privately deployed for telerobotic applications, e.g, multiple users teleoperating multiple robots for some critical tasks in healthcare or industrial applications. For safety, we assume that one user can teleoperate exactly one robot at a time.
%, so that there is a one to one correspondence. 
%Some of these teleoperators can have a line-of-sight of the robot, while some might need a video feedback in real time.
We assume a locally deployed 5G network connects operators and robots, where operators send control commands, and robots transmit real-time video feedback to enable precise decision-making. 

\subsection{Wireless Communication Channel Model}

%\textcolor{blue}{Sabur: will add more text here}
\vspace{-1mm}
We model the 5G wireless network in terms of multiple subcarriers as per the 3GPP standard. The subcarrier resources can be allocated to multiple users in terms of physical resource blocks (PRB). Suppose, there are
%of users in the system at decision time slot $t$ as $n(t)$, and the total number 
$k$ PRBs available in the 5G network. We also assume that the eMBB slice has $n _{e}(t)$ users and URLLC slice has $n_{u}(t)$ users at decision time slot $t$, when the upper-level optimizer will make decision for resource allocation. 
%The number of coexisting eMBB mobile users is m. 
Suppose $p_{ij}(t)$ is the transmit power between the base station gNodeB and user $i$ over PRB $j$ at decision time $t$. 
%For resource allocation in terms of PRBs, 
We define $\rho_{ij}(t)$ as an indicator function such that $\rho_{ij}(t) = 1$, indicates that $PRB_j$ is allocated to user $i$ at decision time slot t; otherwise $\rho_{ij}(t) = 0$. 
Now for wireless channel model, we assume there will be multipath fading as the robots are expected to operate in the indoor or outdoor environments surrounding other structures.
The data rate for each user $i$ is then can be derived as:
\vspace{-2mm}
\begin{equation}\label{eq:datarate}
    r_{i}(t) = B * \sum_{j=1}^{k} \rho_{ij} log_{2}(1 + \frac{p_{ij}(t){h_{ij}}^2(t)}{{\sigma^2}})
\vspace{-1mm}
\end{equation}
where $\sigma^2$ is noise variance, $h_{ij}(t)$ is the time-varying Rayleigh fading channel gain~\cite{mahmud2020performance}, and $B$ is the bandwidth of each PRB which is composed of $12$ subcarriers~\cite{3gpp2017study}.

\subsection{Network Queues}
%In this section, we discuss the formulation of two separate virtual queues for URLLC and eMBB users.
The queue dynamics for the teleop users is influenced by two main factors:  1) The arrival rate into the network depends on the user behavior; 2) The serving (departure) rate is dependent on the robots ability to complete a particular task. 

\vspace{2mm}
\noindent{\bf \textit{Modified Arrival and Departure Rates}}

In teleoperation, the user generates a sequence of set-point commands (e.g., velocity commands) at varying rates depending on user skill, leading to some application delay. Hence, the arrival rates are modeled using a modified Poisson distribution that incorporates delay and bandwidth sensitivity. Let \(\lambda_0\) denote the base arrival rate, \(r_i^u\) the data rate for URLLC users, \(\alpha\) a sensitivity coefficient (\(0 < \alpha < 1\)), and \(D_i\) the application delay. The adjusted arrival rate \(\lambda(u)\) is expressed as:
$\lambda(u) = \lambda_0 - \alpha r_i^u$.
Using this, the probability mass function (PMF) for \(k\) arrivals during a time interval \(\tau + D_i\) is:
\vspace{-2mm}
\begin{equation}\label{pd2}
P_i(k, \tau + D_i) = \frac{((\lambda_0 - \alpha r_i^u)(\tau + D_i))^k e^{-(\lambda_0 - \alpha r_i^u)(\tau + D_i)}}{k!},
\vspace{-1mm}
\end{equation}
where \(D_i\) represents the delay associated with user \(i\).
For departures, we model the serving rate as a function of task efficiency, represented by the Dynamic Task Execution Index (\(\text{DI}\)) defined as:
\begin{equation}\label{di}
\footnotesize
\begin{split}
& \text{DI} = \ 0.4 * \max\left(0, 1 - \frac{\text{\small tracking\_error}}{10}\right) + \\
&  0.3 * \left( \max\left(0, 1 - \frac{\text{\small orientation\_error}}{180}\right) 
+  \max\left(0, 1 - \frac{\text{\small curv}}{1}\right) \right)
\end{split}
\end{equation}
where \(\text{tracking\_error}\), \(\text{orientation\_error}\), and \(\text{curv}\) quantify the robot's performance in tracking set-points, orienting correctly, and minimizing trajectory curvature, respectively.
The serving rate \(\mu_i(DI), \ DI \in (0,1)\) for user \(i\) is modeled as:
\vspace{-2mm}
\begin{equation}\label{sr1}
\mu_i(DI) = r_i^u - \beta \ DI
\vspace{-1mm}
\end{equation}
where $r_i^u$ is the data rate for URLLC robot users and \(\beta > 0\) is a serving coefficient that scales the impact of \(\text{DI}\) on the serving rate. Overall, the URLLC queue, describing the queue of set-point commands, is modeled as:
\vspace{-2mm}
\begin{equation}\label{teleopq1}
F(t+1) = [F(t) + A_1(t) - D(t)]^+
\vspace{-1mm}
\end{equation}
where, $F(t)$: Queue length at time $t$, $A_1(t)= P(t, \tau + D_{max})$, and $D(t)=\mu(DI)$.
For eMBB users, the queue is defined as:
\begin{equation}\label{embbq1}
G(t+1) = [G(t) + A_2(t) - r^e_i]^+
\end{equation}
where \(G(t)\) represents the queue length at time \(t\), \(A_2(t)\) is the arrival rate following a standard Poisson distribution, and \(r_i^e\) is the departure rate for eMBB users.

\section{Proposed Hierarchical Framework}
\begin{figure}[!t]
\centerline{\includegraphics[width=0.99\linewidth]{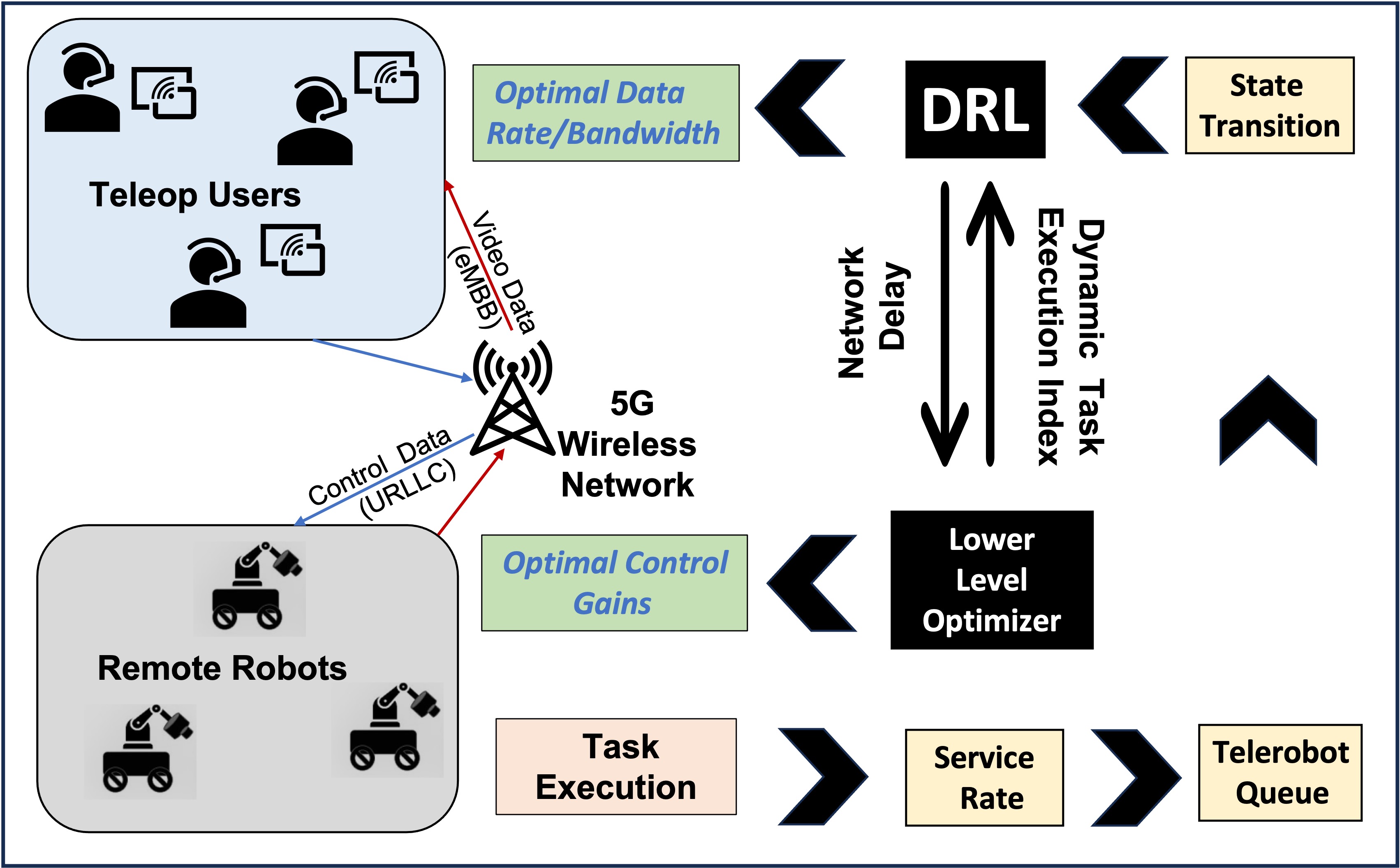}}
\vspace{-1mm}
\caption{\centering \small{Proposed Hierarchical Optimization Framework showing the components and dataflow}}
\label{fig1:framework}
\vspace{-4mm}
\end{figure}

We propose a dual level optimization framework (see fig.~\ref{fig1:framework}) that integrates upper-level network resource allocation with lower-level robot control to address challenges in teleoperation over a shared 5G network. This dual-layered approach ensures efficient resource utilization, network stability, and reliable control for both URLLC and eMBB applications. The upper level uses a Deep Reinforcement Learning (DRL) agent to dynamically allocate resources across network slices, optimizing performance based on real-time conditions and task demands. The DRL agent observes system states, such as virtual queue lengths, and adjusts its policy through rewards that prioritize network efficiency and stability.

At the lower level, robot control optimization leverages the Razumikhin stability criterion to ensure robust performance despite network delays. A matrix inequality computes optimal control gains, minimizing tracking errors and control efforts while maintaining stability under varying network conditions. The upper level resource allocation decisions influence the network delays, which affect the lower level robot control performance. The task execution performance of the robots, in the form of a dexterity index (DI) influence the departure rate and network queue, which in turn updates the DRL agent's state, enabling adaptive policy refinement that aligns network decisions with control outcomes. This interconnected dual level closed-loop design ensures seamless coordination between the network and control systems.

\subsection{Upper Level Optimization Problem}

We optimize the network data rate for telerobotics operations involving $n_e$ eMBB users and $n_u$ URLLC users. Let $D_{ti}$ represent the end-to-end delay for the $i$-th operator-robot pair, with $D_{\text{max}}$ as the URLLC delay deadline. Define a reliability index $\chi_s \in (0, 1]$ representing the percentage of time $D_{ti} < D_{\text{max}}$. 
%We choose a simple quadratic cost function of the form:
%\[{\arg\max_{\rho_{ij}}}\lim_{t \rightarrow \infty} \left( \sum_{\tau = 1}^{t-1} \sum_{i=1}^{n_u}  (r_u^i(\tau))^2 +\sum_{\tau = 1}^{t-1}\sum_{i=1}^{n_e}  (r_e^i(\tau))^2\right).\]
This Optimization problem can be formulated as a minimization problem as follows:
\begin{equation}\label{oppf1}
\footnotesize{
{\arg\min_{\rho_{ij}}}\lim_{t \rightarrow \infty} \left( \sum_{\tau = 1}^{t-1} \sum_{i=1}^{n_u}  \frac{1}{(r_i^u)^2 (\tau)+\epsilon }+\sum_{\tau = 1}^{t-1}\sum_{i=1}^{n_e}  \frac{1}{(r_i^e)^2(\tau)+\epsilon}\right)}
\end{equation}
subject to constraints:
\setlength{\jot}{0pt} 
\begin{subequations}\label{constr1}
\footnotesize
\begin{align}
\label{3a}
& \text{Reliability:}
P (D_{ti} > D_{max}) \leq 1 - \chi_s, \ \forall i = 1, \hdots, n_u, \\ 
\label{3b}
& \text{Resource Allocation:}
\sum_{j=1}^{k} \sum_{i=1}^{n_u + n_e} \rho_{ij} = K, \\
\label{3c}
& \sum_{i=1}^{n_u + n_e } \rho_{ij} = 1 \quad \forall \; j = 1, 2, \ldots, k, \\
\label{3d}
& \sum_{j=1}^{K} \rho_{ij} \geq 1 \quad \forall \; i = 1, 2, \ldots, n_u + n_e 
\end{align}
\end{subequations}
where $r_i^u$ and  $r_i^e$ are the URLLC and eMBB data rates respectively, both of which are function of $\rho_{ij}$ as mentioned in equation~\eqref{eq:datarate}. The parameter $\epsilon$ is a small positive scalar, which is introduced to keep the minimization problem computationally feasible in an unwanted case of very low data rate.\\
To satisfy Lyapunov stability, define virtual queues for URLLC ($F(k)$) and eMBB ($G(k)$):
\begin{subequations}\label{dynamic}
\begin{align}
\label {6a}
&F(k+1) = F(k) + A_1(k) - D(k)
\\ \label{6b} 
& G(k+1) = G(k) + A_2(k) - r_i^e
\end{align}
\end{subequations}
Using a candidate network Lyapunov function:
\vspace{-2mm}
\begin{equation}\label{eq:lyapunov_function}
L(F(k), G(k)) = \frac{1}{2} \left(F(k)^2 + G(k)^2 \right),
\vspace{-1mm}
\end{equation}
the Lyapunov drift-plus-penalty equation becomes:
\vspace{-2mm}
\begin{equation}\label{eq:drift_penalty}
\Delta L(k) + V h(r_i),
\vspace{-1mm}
\end{equation}
where $h(r_i)$ is the original cost function, and $V > 0$ is a penalty weight. The delay constraint can be reformulated as:
\vspace{-2mm}
\begin{equation}\label{eq:delay_constraint}
y(\tau) = e^{-\alpha_1 r_i^u(\tau) D_{\text{max}}} - (1 - \chi_s) \geq 0,
\vspace{-1mm}
\end{equation}
where $\alpha_1 > 0$ determines the impact of data rate on delay.

\vspace{2mm}
\noindent{\bf \textit{Lagrangian Formulation}}

We combine the cost function and the constraint through a Lagrangian function. The Lagrangian for the unconstrained problem is chosen as:
\vspace{-2mm}
\begin{equation}\label{lagrang1}
L_a(r_i, y(\tau), \lambda_l) = \Delta L(\tau) + V h(r_i) - \lambda_l y(\tau),
\vspace{-1mm}
\end{equation}
where $\lambda_l \geq 0$ is the Lagrange multiplier. The optimization becomes:
\vspace{-2mm}
\begin{equation}\label{unconstp1}
\arg\min_{\rho_{ij}} \arg\max_{\lambda_l > 0} L_a(r_i, y(\tau), \lambda_l).
\vspace{-1mm}
\end{equation}

This unified framework incorporates reliability, resource allocation, and delay constraints into a single optimization problem.

\subsection{Lower Level Optimization}

We model the robot dynamics to be a discrete-time linear system with delay \(T_d\) as:  
\vspace{-2mm}
\begin{equation}\label{sysdyn2}
x(k+1) = A x(k) + A_d x(k - T_d) + B u(k)
\vspace{-1mm}
\end{equation}  
where the states are represented as $x(k)$ with system matrices $A, A_d$ and $B$. With a state feedback control law \(u(k) = -K x(k)\), the closed loop dynamics becomes: 
\vspace{-2mm}
\[
x(k+1) = (A - B K) x(k) + A_d x(k - T_d).
\vspace{-1mm}
\]  
For a candidate Lyapunov function \(V(x(k)) = x(k)^T P x(k)\), we can exploit the Razumikhin's stability criterion:  
\vspace{-2mm}
\[
V(x(k - T_d)) \leq \gamma V(x(k)), \quad \Delta V(x(k)) \leq -\alpha V(x(k)),
\vspace{-1mm}
\]  
where \(\alpha > 0\) and \(\gamma = 1 + \alpha T_d\), for ensuring stability.

Using the augmented state vector \(\zeta(k) = \begin{bmatrix} x(k) \\ x(k - T_d) \end{bmatrix}\), the stability condition can be derived as a matrix inequality:  
\begin{equation}\label{modlmi}
\footnotesize
\begin{bmatrix} 
A^T P A - P + \alpha P & A^T P A_d & A^T P B Y \\ 
A_d^T P A & \gamma A_d^T P A_d & A_d^T P B Y \\ 
Y^T B^T P A & Y^T B^T P A_d & Y^T B^T P B Y - P 
\end{bmatrix} < 0,
\end{equation}
where \(Y = K P\), \(\gamma = 1 + \alpha T_d^{\max}\), and \(T_d^{\max}\) is the maximum anticipated delay. This ensures stability under nominal scenario provided the sampling period exceeds the network delay. For dealing with the stochastic condition for delay \(T_d\), a convex optimization problem minimizing \(\| \hat{K} - K \|^2\) need to be solved along with \eqref{modlmi} as a constraint, where \(\hat{Y} = \hat{K} P\) points to the adjusted gain $\hat{K}$.

%\vspace{4mm}
\subsection{ Deep Reinforcement Learning (DRL) Framework }

%The problem of resource allocation in this scenario involves efficiently distributing resources to meet the demands of both eMBB (enhanced Mobile Broadband) users and URLLC (Ultra-Reliable Low-Latency Communication) users. 
The goal is to minimize the average cost while satisfying the constraints \(\ref{3b}\) and \(\ref{3c}\). This problem is framed as a Markov Decision Process (MDP), where decisions are made under uncertainty. To solve this, the Advantage Actor-Critic (A2C) algorithm, a Deep Reinforcement Learning (DRL) method, is applied. A2C integrates policy optimization with value estimation, stabilizing the learning process. In this framework, the Actor learns a policy that outputs action probabilities, while the Critic estimates the expected future rewards. The advantage function refines the Actor’s decisions by providing feedback on whether actions lead to better or worse outcomes than expected, thus improving the learning process.

In our setup, two independent agents handle resource allocation: one for eMBB users and another for URLLC users. These agents aim to optimally allocate \(K\) physical resource blocks (PRBs) while satisfying system constraints. The parameters used in the pseudocode (Algorithm~\ref{algo}) include the Actor and Critic networks for both agents, represented by \(\theta_{a1}\), \(\theta_{c1}\) for Agent 1 (eMBB) and \(\theta_{a2}\), \(\theta_{c2}\) for Agent 2 (URLLC). Each agent uses a replay buffer (\(\text{Replay buffer1}\), \(\text{Replay buffer2}\)) to store experiences and update their networks. The learning rate and discount factor \(\gamma\) control the learning process, with each agent updating their Actor and Critic networks at each time step. The reward function \(r_1\) and \(r_2\) guide the agents to minimize penalties, while the Lagrangian multiplier \(\lambda\) is used to adjust the constraints, ensuring the optimal allocation strategy is both effective and compliant with system requirements. As outlined in the pseudocode, at each time step, the agents interact with the environment, select actions \(a_1\) and \(a_2\) based on their policies \(\pi(s_1 | \theta_{a1})\) and \(\pi(s_2 | \theta_{a2})\), respectively, and update their parameters using the calculated advantages.

\setlength{\textfloatsep}{0pt}
\begin{algorithm}[!t]
\footnotesize
\caption{Dual-Agent DDPG for Resource Allocation}
\begin{algorithmic}[1]
    \STATE \textbf{Initialize:} Actor and Critic networks for each agent with parameters $\theta_{a1}, \theta_{c1}, \theta_{a2}, \theta_{c2}$; Replay buffers for each agent; Set hyperparameters

    \FOR{each episode}
        \STATE Reset environment and observe initial states $s^1$, $s^2$

        \FOR{each time step}
            \STATE \textbf{Agent 1:} Select action $a^1$ with Actor1, execute it, receive $r^1$, $s^{1'}$
            \STATE \textbf{Agent 2:} Select action $a_2$ with Actor2, execute it, receive $r^2$, $s^{2}$
            \STATE Store transitions $(s^1, a^1, r^1, s^{1'})$ in Replay buffer1, $(s^2, a^2, r^2, s^{2'})$ in Replay buffer2
            \STATE \textbf{Update Agent1:} Sample batch from Replay buffer1, calculate target $y^1$ and advantage, update Critic1 and Actor1
            \STATE \textbf{Update Agent2:} Sample batch from Replay buffer2, calculate target $y^2$ and advantage, update Critic2, $\lambda$, and Actor2
            \STATE Update states: $s^1 \leftarrow s^{1'}$, $s^2 \leftarrow s^{2'}$
        \ENDFOR
    \ENDFOR
\end{algorithmic}
\label{algo}
\end{algorithm}
%The Actor-Critic method shown in Figure \ref{fig:actor_critic} is a powerful reinforcement learning approach that combines the strengths of policy-based and value-based methods. It consists of two main components: the actor and the critic. The actor is responsible for selecting actions based on the current state, effectively learning and improving the policy that dictates the action choices. Meanwhile, the critic evaluates the actions taken by the actor by estimating the value function, which represents the expected cumulative reward from a given state. The critic provides feedback to the actor by calculating the temporal difference (TD) error, which measures the discrepancy between the predicted value and the actual reward received. This feedback helps the actor refine its policy to make better decisions over time. By iterating through this process, the Actor-Critic method allows the agent to learn an optimal policy that maximizes long-term rewards, making it particularly well-suited for solving complex decision-making problems like the resource allocation problem in an MDP framework.

%\begin{figure} [h]
%\vspace{-2mm}
%\centerline{\includegraphics[width=0.9\linewidth, height = 10cm]{images/picture3.png}}
%\vspace{-1mm}
%\caption{DRL Framework For Solving Optimization Problem}
%\label{fig:actor_critic}
%\vspace{2mm}
%\end{figure}

\vspace{-1mm}
\section{Experiment and Results}

\subsection{Simulation Setup}
We implemented our DRL framework in Python and simulated it using realistic 5G parameters (Table I). The 5G network has a 400 m radius with two slices: eMBB and URLLC. Network conditions follow a defined channel model, and packet arrivals follow a Poisson distribution. Practical values for control dexterity, URLLC delay, and reliability constraints were considered. We assume user counts do not exceed available PRBs, with a one-to-one mapping between telerobotic users and robots, and users of each type (URLLC or eMBB) are independently and identically distributed.
%We  test our proposed DRL framework based on different scenarios, which arise when the network dynamics changes due to a corresponding variable of interest. 
%The DRL 

\begin{center}
\footnotesize
\centering
\begin{table}[!b]
\vspace{1mm}
    \centering
    \begin{tabular}{ |c|c|c| } 
 \hline
\textbf{Parameter}&\textbf{Symbols} & \textbf{Values}  \\ 
 \hline
 Bandwidth &$B$ & 10 MHz  \\ 
 \hline
 Noise Variance & $\sigma^2$ & -110 dBm \\
 \hline
  Reliability & $\chi_s$ & $95\%$ \\
 \hline
 Deadline for URLLC Delay& $D_{\max}$ & 20 ms\\
 \hline
 Transmit power& $p_{ij}$ & 20 dBm\\
 \hline
 Number of PRBs & K & 25\\
 \hline
 modified Poisson distribution alpha & $\alpha$  &0.1 \\
 
 \hline
\end{tabular}
\vspace{-1mm}
\caption{\centering \small{Network Simulation Parameters}}
\label{tab1:specificatio}
\vspace{-3mm}
\end{table}
\end{center}

\iffalse
\begin{figure}[!t]
    \centering
    \includegraphics[width=0.6\linewidth]{images/total_reward.png}
    \vspace{-2mm}
    \caption{Total Reward over different iterations in DRL}
    \label{fig:total_reward}
\end{figure}
\fi

%\begin{figure}[h]
%    \centering
%    \includegraphics[width=0.5\linewidth]{images/embb_reward.png}
%    \caption{eMBB Reward}
%    \label{fig:embb_reward}
%\end{figure}

%\begin{figure}[h]
%    \centering
%    \includegraphics[width=0.5\linewidth]{images/urllc_reward.png}
%    \caption{URLLC Reward}
 %   \label{fig:embb_reward}
%\end{figure}

\begin{figure}[!t]
\vspace{-3mm}
    \centering
    {\begin{subfigure}[]{0.49\linewidth}
        \centering
        \includegraphics[width=0.99\textwidth]{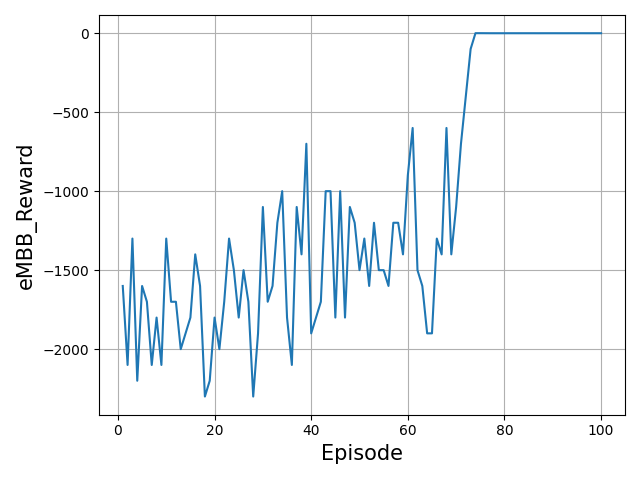}
        \vspace{-6mm}
        \caption{eMBB Reward}
        \label{fig:embb_reward}
    \end{subfigure}
    \hfill
    \begin{subfigure}[]{0.49\linewidth}
        \centering
        \includegraphics[width=0.99\textwidth]{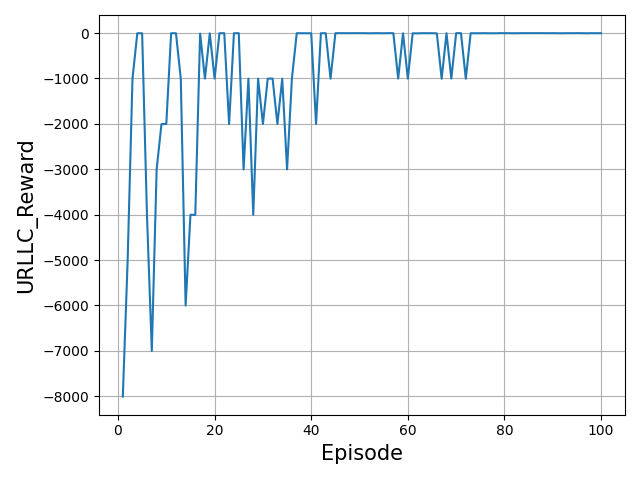}
        \vspace{-6mm}
        \caption{URLLC Reward}
        \label{fig:urllc_reward}
    \end{subfigure}}   
    \vspace{-2mm}
    \caption{\small{Reward for eMBB and URLLC Slices}}
    \label{fig:slice_reward}
    \vspace{-1mm}
\end{figure}

%\begin{figure}[h]
%    \centering
%    \includegraphics[width=0.5\linewidth]{images/embb_ql.png}
%    \caption{eMBB Queue Length}
%    \label{fig:embb_ql}
%\end{figure}

%\begin{figure}[h]
%    \centering
%    \includegraphics[width=0.5\linewidth]{images/urllc_ql.png}
%    \caption{URLLC Queue Length}
%    \label{fig:embb_reward}
%\end{figure}

\begin{figure}[!t]
%\vspace{-2mm}
    \centering
    {\begin{subfigure}[]{0.49\linewidth}
        \centering
        \includegraphics[width=0.99\textwidth]{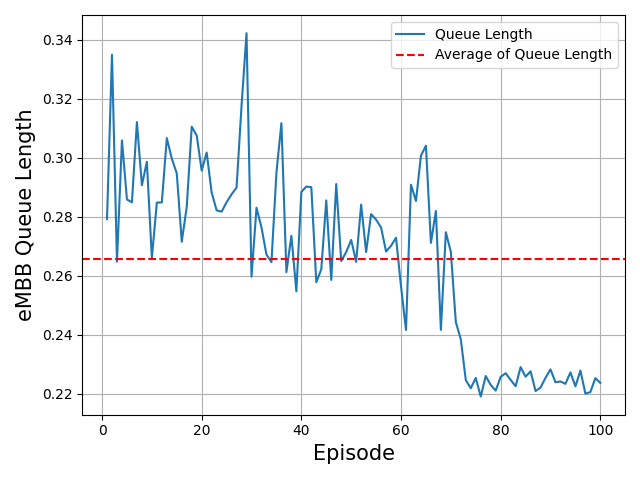}
        \vspace{-6mm}
        \caption{eMBB Queue Length}
        \label{fig:embb_ql}
    \end{subfigure}
    \hfill
    \begin{subfigure}[]{0.49\linewidth}
        \centering
        \includegraphics[width=0.99\textwidth]{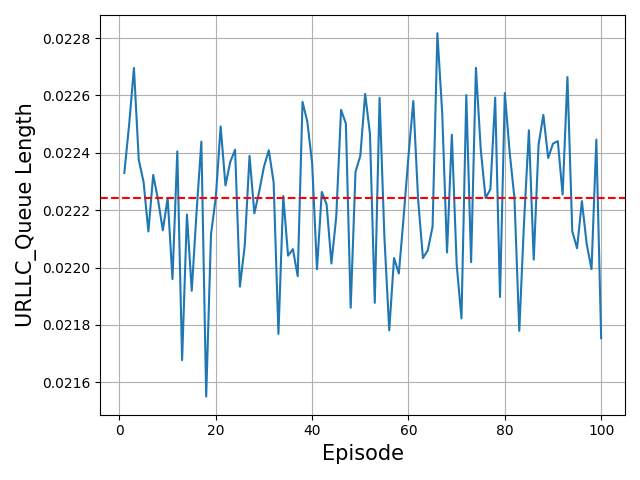}
        \vspace{-6mm}
        \caption{URLLC Queue Length}
        \label{fig:urllc_ql}
    \end{subfigure}}   
    \vspace{-2mm}
    \caption{\small{Queue length of eMBB and URLLC users}}
    \label{fig:q_len}
    \vspace{-4mm}
\end{figure}

\vspace{-8mm}
\subsection{Results}

\vspace{-1mm}
\subsubsection{\textbf{DRL Performance}}
We measure the performance of DRL that solves the upper-level optimization. The total reward is converged, and this reward is the sum of the eMBB and URLLC rewards as shown in~\cref{fig:embb_reward,fig:urllc_reward}.
Furthermore, the queue length for the users of those slices is displayed in~\cref{fig:embb_ql,fig:urllc_ql}. There has been a reduction in the queue length during training in the final episodes of the series for eMBB users.

\subsubsection{\textbf{Varying Network Conditions}}
\begin{figure}[!t]
%\vspace{-6mm}
    \centering
    {\begin{subfigure}{0.49\linewidth}
        \centering
        \includegraphics[width=0.95\textwidth]{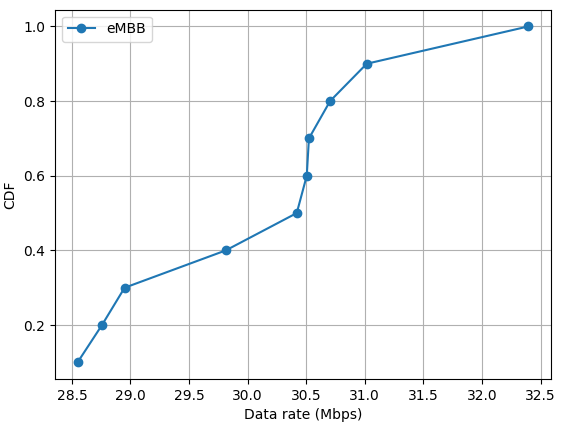}
        %\vspace{-4em}
        \caption{CDF of eMBB datarate}
        \label{fig:embb_cdf}
    \end{subfigure}
    \hfill
    \begin{subfigure}
    {0.49\linewidth}
    %\vspace{4mm}
        \centering
        \includegraphics[width=0.95\textwidth]{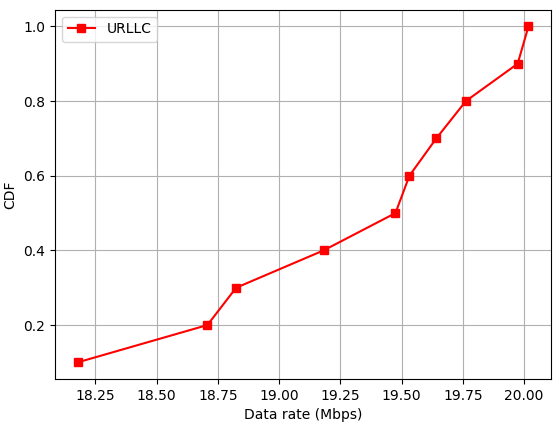}
        %\vspace{-4em}
        \caption{CDF of URLLC datarate}
        \label{fig:urllc_cdf}
    \end{subfigure}}   
    \vspace{-6mm}
    \caption{\small{CDF of eMBB and URLLC datarates with Varying $\sigma$ (Rayleigh Fading Scale) in the Wireless Network.}}
    \label{fig:datarate_cdf}
    \vspace{1mm}
\end{figure}

%As we modeled that our wireless channel is impacted by multipath fading with Rayleigh distribution,
We measure the data rate over the time varying wireless channel conditions.  Figs.~\ref{fig:embb_cdf} and \ref{fig:urllc_cdf} show the cumulative distribution function (CDF) of the data rate for the eMBB and URLLC users respectively. As we allocated more bandwidth to the eMBB slice, eMBB users get higher bandwidth as expected. For this experiment we chose the scaling factor of the Rayleigh fading as $1.0$. The CDF shows the throughput are more or less evenly distributed in the lower and upper region with a sharper slope in the middle, indicating the mid-datarate value achieved frequently.

%\subsubsection{Varying Channel Condition}
%\begin{figure}[h]
%    \centering
%    \includegraphics[width=0.5\linewidth]{images/h_var.png}
%    \caption{Varying Channel Condition ($h_{ij}$)}
%    \label{fig:h_var}
%\end{figure}

\iffalse

\begin{figure}[!t]
\vspace{-3mm}
    \centering   \includegraphics[width=0.8\linewidth]{images/var_h_datarate.png}
    \vspace{-2mm}
    \caption{Avg. eMBB and URLLC Data Rate for Varying $\sigma$ (Rayleigh Fading Scale) in the Wireless Network}
    \label{fig:var_h_datarate}
\end{figure}

\begin{figure}[!t]
\vspace{-3mm}
    \centering \includegraphics[width=0.8\linewidth]{images/var_h_ql.png}
    \vspace{-2mm}
    \caption{Avg. eMBB and URLLC Queue Length for Varying $\sigma$ (Rayleigh Fading Scale) in the Wireless Network}
    \label{fig:var_h_ql}
\end{figure}

\fi

We also vary the Rayleigh fading distribution scaling factor which can potentially indicate varying multipath conditions in indoor, outdoor and cluttered environments. Fig.~\ref{fig:var_h_datarate} shows the avg. data rate of the eMBB and URLLC users for increasing scaling factor which reducing  peaks of the Rayleigh distribution indicating better channel conditions along X-axis. Both eMBB and URLLC data rate shows improvement due to better channel conditions, without any change in the bandwidth allocation policy. Fig.~\ref{fig:var_h_ql} shows the avg. queue length of the eMBB and URLLC users for varying Rayleigh fading scaling factor. It shows that with improved network conditions, i.e., higher scaling factor, as the network can carry more data, the queue length also slowly increases for the eMBB users, but still the queue length is finite and not very high. For URLLC users, the queue length is much smaller due to smaller control data packets and, also better optimized control gain by the lower-level optimizer. It remains almost unaffected with the network variations.

\subsubsection{\textbf{Varying Sampling Interval}}
We wanted to check the effects of varying sampling intervals in the lower level controllers. The optimal data rate resulted from these variations are shown in Fig.~\ref{fig:varying_sampling}. The minor variations in the eMBB and URLLC data rate observed in Fig.~\ref{fig:varying_sampling} with changes in the sampling interval demonstrate the robustness of our proposed resource allocation scheme.  The upper level DRL agent effectively adapts the resource allocation to maintain a consistent data rate for both users, even when the sampling period of the robot control system changes. The changes in sampling period affect the robustness of the nominal gains and may potentially degrade task performance. However, the upper level adapt to the lower level task execution by dynamically changing the data rate according to the situation. 
%This adaptability is crucial for real-time teleoperation applications where control system parameters (like sampling period) might need to be adjusted. 
%\begin{figure}[h]
%    \centering
%    \includegraphics[width=0.8\linewidth]{images/samplin_var.png}
%    \caption{Average Data rate for eMBB and URLLC users with different sampling interval control}
%    \label{fig:sampling_var}
%\end{figure}

\begin{figure}[!t]
%\vspace{-3mm}
    \centering
    {\begin{subfigure}{0.49\linewidth}
        \centering
        \includegraphics[width=0.99\textwidth]{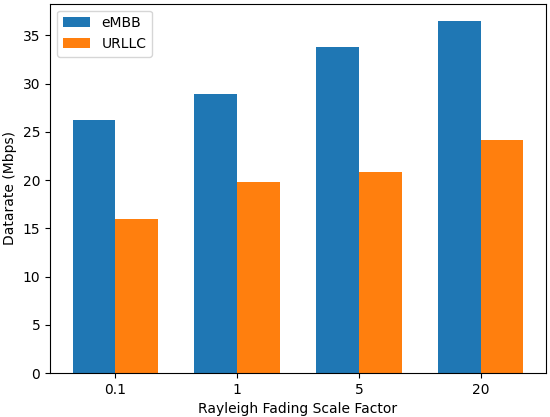}
        %\vspace{-4em}
        \caption{Datarate vs varying $\sigma$}
        \label{fig:var_h_datarate}
    \end{subfigure}
    \hfill
    \begin{subfigure}
    {0.49\linewidth}
    %\vspace{4mm}
        \centering
        \includegraphics[width=0.99\textwidth]{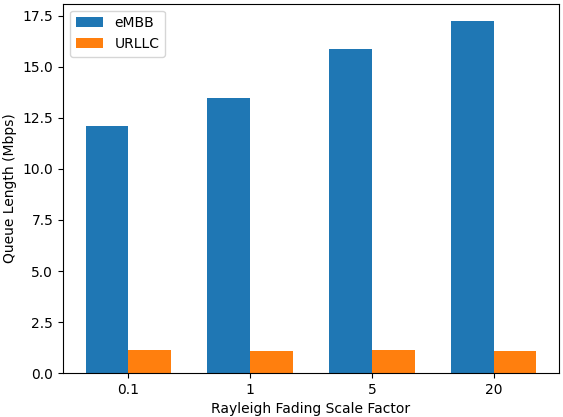}
        %\vspace{-4em}
        \caption{Queue length vs $\sigma$}
        \label{fig:var_h_ql}
    \end{subfigure}}   
    \vspace{-6mm}
    \caption{\small{Avg. eMBB and URLLC Datarate and Queue Length for Varying $\sigma$ (Rayleigh Fading Scale) in the Wireless Network}}
    \label{fig:datarate_cdf}
    \vspace{-4mm}
\end{figure}

\begin{figure}[!t]
%\vspace{-2mm}
    \centering
    {\begin{subfigure}[]{0.49\linewidth}
        \centering
        \includegraphics[width=0.97\textwidth]{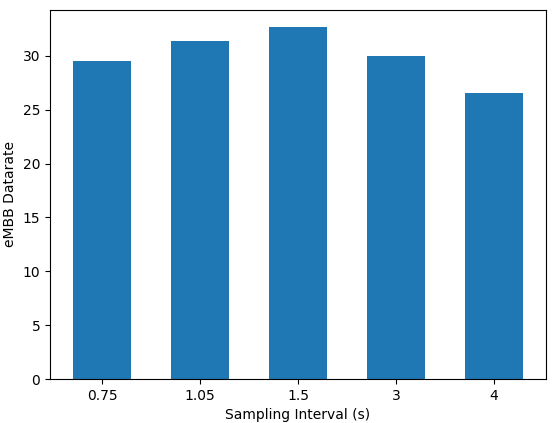}
        %\vspace{-4em}
        \caption{eMBB datarate vs\\ Sampling interval}
        \label{fig:embb_sm}
    \end{subfigure}
    \hfill
    \begin{subfigure}[]{0.49\linewidth}
        \centering
        \includegraphics[width=0.97\textwidth]{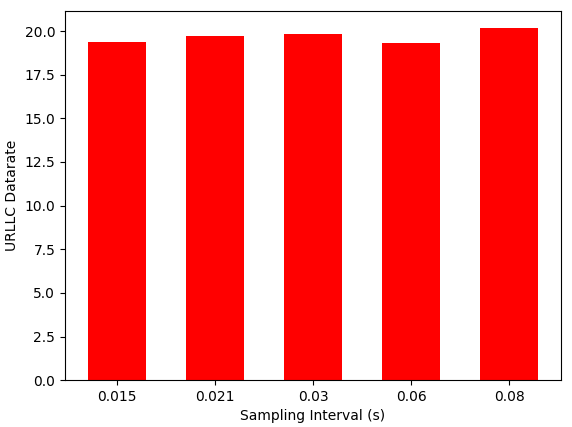}
        %\vspace{-4em}
        \caption{URLLC datarate vs\\ Sampling interval}
        \label{fig:urllc_sm}
    \end{subfigure}}   
    \vspace{-2mm}
    \caption{\small{Average Data rate for eMBB and URLLC users with different sampling interval control}}
    \label{fig:varying_sampling}
    %\vspace{-2mm}
\end{figure}

\vspace{1mm}
\subsubsection{\textbf{Varying DI}}
\begin{figure}[!t]
%\vspace{-2mm}
    \centering
    {\begin{subfigure}[]{0.49\linewidth}
        \centering
        \includegraphics[width=0.95\textwidth]{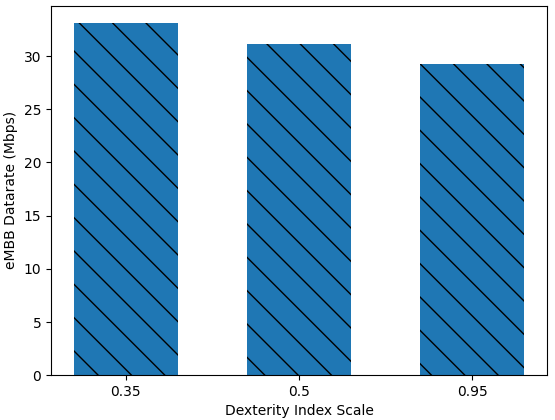}
        \vspace{-2mm}
        \caption{eMBB datarate vs DI}
        \label{fig:embb_di}
    \end{subfigure}
    \hfill
    \begin{subfigure}[]{0.49\linewidth}
        \centering
        \includegraphics[width=0.95\textwidth]{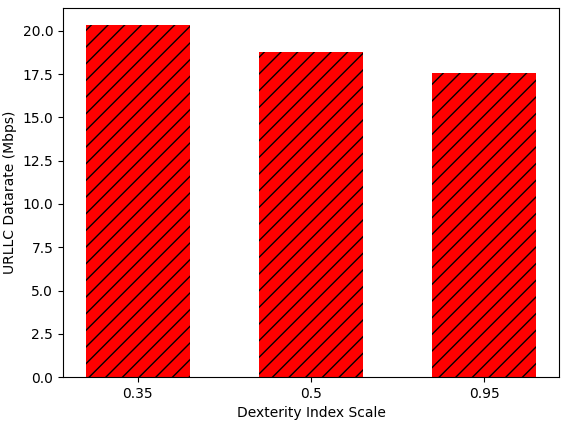}
        \vspace{-2mm}
        \caption{URLLC datarate vs DI }
        \label{fig:urllc_di}
    \end{subfigure}}   
    \vspace{-2mm}
    \caption{\small{Average Data rate for eMBB and URLLC users with different Dynamic Task Execution Index values}}
    \label{fig:varying_DI}
    \vspace{-4mm}
\end{figure}

To examine the effects of task execution variation on the optimal data rate, we choose 3 different set of reference signals which gave us low, moderate and high DI (through trial and error). As shown in the Fig.~\ref{fig:varying_DI}, lower DI results in a higher data rate and higher DI produces lower data rate. As the dexterity index increases, indicating more complex and demanding tasks, the DRL agent intelligently adjusts the resource allocation, potentially lowering the data rate to provide the robots with more time for task execution, effectively balancing performance and efficiency. This adaptability is crucial for supporting a wide range of teleoperation tasks with varying dexterity requirements. The dynamic variations in the URLLC and eMBB data rate with changes in the DI (Fig.~\ref{fig:varying_DI}) highlight the adaptability of our DRL-based resource allocation to the varying demands of teleoperation task execution.
%\begin{figure}[h]
%    \centering
%    \includegraphics[width=0.5\linewidth]{images/di_var.png}
%    \caption{Average Data rate for eMBB and URLLC users with different Dexterity Index Scale}
%    \label{fig:di_var}
%\end{figure}

\begin{figure}[!t]
%\vspace{-3mm}
    \centering   \includegraphics[width=0.7\linewidth]{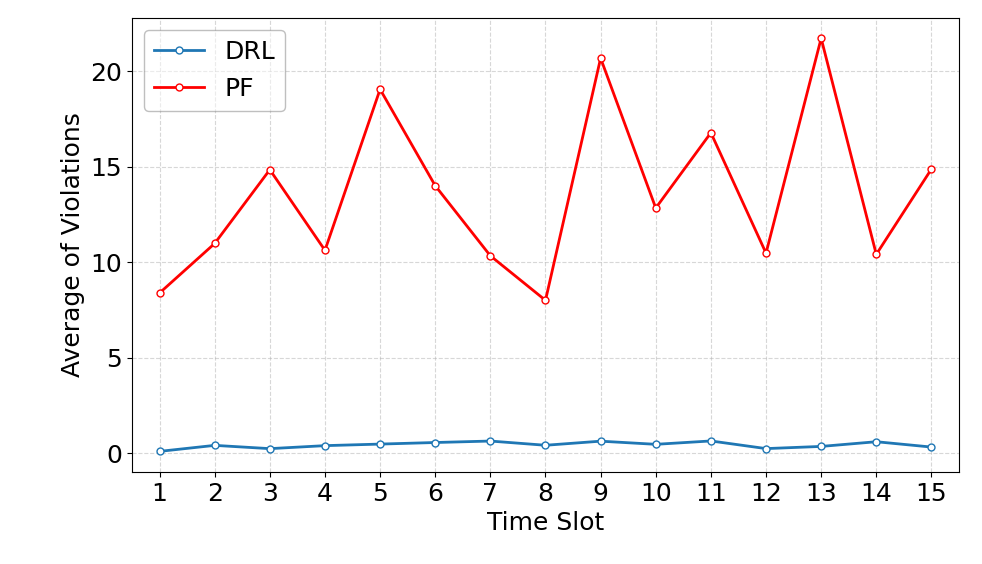}
    \vspace{-4mm}
    \caption{\small{Average of Delay Violation  Across all URLLC users for our method (DRL) and Proportional Fair (PF) allocation}}
    \label{fig:violation_comparison}
    \vspace{-4mm}
\end{figure}

\vspace{1mm}
\subsubsection{\textbf{Adjusted Gain}}
We wanted to compare the nominal gains and adjusted gains when the delay constraint is violated. The results in table~\ref{table:tr_comparison} compares the tracking error of the robot control system for two different control gain settings. The table shows that the adjusted gain generally leads to a lower tracking error compared to the nominal gain when the delay constraint is violated. This highlights the effectiveness of the proposed gain adjustment in maintaining satisfactory task execution even when the network quality degrades.

\subsubsection{\textbf{Comparison}}
Proportional Fair (PF) resource allocation is a widely used method in wireless networks that balances throughput and fairness by maximizing the sum of the logarithms of users' data rates. We compared PF based resource allocation with our proposed method. While the datarate is comparable in both (table~\ref{tab:dr_comparison}), the delay violations are significantly lower with the proposed approach as shown in in figure~\ref{fig:violation_comparison}. This highlights its effectiveness in meeting stringent latency requirements, particularly for applications such as telerobotics, where minimizing delay violations is critical for reliable and responsive operation.

\begin{table}[!t]
\vspace{4mm}
    \centering
    \begin{tabular}{|c|c|c|}
        \hline
        \multirow{2}{*}{\shortstack{\textbf{Delay Violation}\\ \textbf{Time Slot}}} & \multicolumn{2}{c|}{\textbf{Tracking Error}} \\
        \cline{2-3}
        & \textbf{Nominal Gain} & \textbf{Adjusted Gain} \\
        \hline
        5 & 0.8962 & 0.7077 \\
        \hline
        10 & 0.7877 & 0.5609 \\
        \hline
        
        20 & 0.8726 & 0.5660 \\
        \hline
    \end{tabular}
    \vspace{-2mm}
    \caption{\small{Comparison of Tracking Error for Nominal and Adjusted Gain in Delay Violation Time Slots}}
    \label{table:tr_comparison}
\end{table}

\begin{table}[!t]
\vspace{-4mm}
    \centering
\begin{tabular}{|c|c|c|c|}
\cline { 3 - 4 } \multicolumn{2}{c|}{} & \multicolumn{2}{c|}{ \textbf{Datarate} (Mbps) } \\
\hline \textbf{Timeslot} & \textbf{Channel gain} & \textbf{DRL} & \textbf{PF} \\
\hline 
5& 1.0481&17.9188 & 19.4697\\
\hline
10&1.0645 &19.6458 & 19.2875\\
\hline 
15&1.8141 &19.7586 &19.9794\\
\hline
\end{tabular}
\vspace{-2mm}
    \caption{\small{Comparison of Data rate for DRL and Proportional fair allocation, Average across all URLLC users}}
    \label{tab:dr_comparison}
    %\vspace{-1mm}
\end{table}

%\subsection{Discussions}
\vspace{-1mm}
\section{Conclusions}
This paper presents a novel hierarchical optimization framework for resource allocation in 5G networks serving a mix of URLLC and eMBB applications, with a particular focus on telerobotics.  We address the challenges of varying robotic task execution requirements in teleoperation by integrating the upper level DRL with a lower-level Razumikhin-based controller that adapts to network delays, ensuring stability and performance. 
%The upper-level DRL agent dynamically allocates resources across different users, while maintaining overall network stability. Our dual-queue model captures the distinct needs of URLLC-driven telerobotic control and eMBB data traffic, and the hierarchical structure allows for efficient adaptation to fluctuating network conditions and task demands.  
Simulation results demonstrate the efficacy of our approach in achieving robust performance and efficient resource utilization across diverse scenarios, highlighting its potential for enabling reliable and responsive teleoperation in the evolving 5G landscape. Future work will address generalization to unknown robot dynamics and uncertain network conditions.

\balance
\vspace{-2mm}
\bibliographystyle{IEEEtran}
\bibliography{ref}

\begin{thebibliography}{10}
\providecommand{\url}[1]{#1}
\csname url@samestyle\endcsname
\providecommand{\newblock}{\relax}
\providecommand{\bibinfo}[2]{#2}
\providecommand{\BIBentrySTDinterwordspacing}{\spaceskip=0pt\relax}
\providecommand{\BIBentryALTinterwordstretchfactor}{4}
\providecommand{\BIBentryALTinterwordspacing}{\spaceskip=\fontdimen2\font plus
\BIBentryALTinterwordstretchfactor\fontdimen3\font minus \fontdimen4\font\relax}
\providecommand{\BIBforeignlanguage}[2]{{%
\expandafter\ifx\csname l@#1\endcsname\relax
\typeout{** WARNING: IEEEtran.bst: No hyphenation pattern has been}%
\typeout{** loaded for the language `#1'. Using the pattern for}%
\typeout{** the default language instead.}%
\else
\language=\csname l@#1\endcsname
\fi
#2}}
\providecommand{\BIBdecl}{\relax}
\BIBdecl

\bibitem{leng2022industry}
J.~Leng, W.~Sha, B.~Wang, P.~Zheng, C.~Zhuang, Q.~Liu, T.~Wuest, D.~Mourtzis, and L.~Wang, ``Industry 5.0: Prospect and retrospect,'' \emph{Journal of Manufacturing Systems}, vol.~65, pp. 279--295, 2022.

\bibitem{demir2019industry}
K.~A. Demir, G.~D{\"o}ven, and B.~Sezen, ``Industry 5.0 and human-robot co-working,'' \emph{Procedia computer science}, vol. 158, pp. 688--695, 2019.

\bibitem{khalid2016methodology}
A.~Khalid, P.~Kirisci, Z.~Ghrairi, K.-D. Thoben, and J.~Pannek, ``A methodology to develop collaborative robotic cyber physical systems for production environments,'' \emph{Logistics Research}, vol.~9, pp. 1--15, 2016.

\bibitem{choi2018telesurgery}
P.~J. Choi, R.~J. Oskouian, and R.~S. Tubbs, ``Telesurgery: past, present, and future,'' \emph{Cureus}, vol.~10, no.~5, 2018.

\bibitem{khan2022urllc}
B.~S. Khan, S.~Jangsher, A.~Ahmed, and A.~Al-Dweik, ``Urllc and embb in 5g industrial iot: A survey,'' \emph{IEEE Open Journal of the Communications Society}, vol.~3, pp. 1134--1163, 2022.

\bibitem{ali2021urllc}
R.~Ali, Y.~B. Zikria, A.~K. Bashir, S.~Garg, and H.~S. Kim, ``Urllc for 5g and beyond: Requirements, enabling incumbent technologies and network intelligence,'' \emph{IEEE Access}, vol.~9, pp. 67\,064--67\,095, 2021.

\bibitem{rayguru2020}
M.~M. Rayguru, M.~R. Elara, B.~F. Gómez, and B.~Ramalingam, ``A time delay estimation based adaptive sliding mode strategy for hybrid impedance control,'' \emph{IEEE Access}, vol.~8, pp. 155\,352--155\,361, 2020.

\bibitem{10.23919/icn.2022.0011}
X.~Han, K.~Xiao, R.~Liu, X.~Liu, and G.~Alexandropoulos, ``Dynamic resource allocation schemes for embb and urllc services in 5g wireless networks,'' \emph{Intelligent and Converged Networks}, 2022.

\bibitem{10.1109/camad.2019.8858433}
P.~Korrai, E.~Lagunas, S.~Sharma, S.~Chatzinotas, and B.~Ottersten, ``Slicing based resource allocation for multiplexing of embb and urllc services in 5g wireless networks,'' 2019.

\bibitem{10.1109/mcom.2018.1701178}
H.~Chen, R.~Abbas, P.~Cheng, M.~Shirvanimoghaddam, W.~Hardjawana, W.~Bao, Y.~Li, and B.~Vucetic, ``Ultra-reliable low latency cellular networks: use cases, challenges and approaches,'' \emph{Ieee Communications Magazine}, vol.~56, pp. 119--125, 2018.

\bibitem{10.1109/access.2018.2872781}
P.~Popovski, K.~F. Trillingsgaard, O.~Simeone, and G.~Durisi, ``5g wireless network slicing for embb, urllc, and mmtc: a communication-theoretic view,'' \emph{IEEE Access}, vol.~6, pp. 55\,765--55\,779, 2018.

\bibitem{10.1109/lcomm.2019.2959335}
E.~Santos, R.~Souza, J.~Rebelatto, and H.~Alves, ``Network slicing for urllc and embb with max-matching diversity channel allocation,'' \emph{Ieee Communications Letters}, vol.~24, pp. 658--661, 2020.

\bibitem{10.3390/math11183925}
E.~Makeeva, I.~Kochetkova, and R.~Alkanhel, ``Retrial queueing system for analyzing impact of priority ultra-reliable low-latency communication transmission on enhanced mobile broadband quality of service degradation in 5g networks,'' \emph{Mathematics}, vol.~11, p. 3925, 2023.

\bibitem{10.1109/lwc.2020.3046628}
W.~Zhang, M.~Derakhshani, and S.~Lambotharan, ``Stochastic optimization of urllc-embb joint scheduling with queuing mechanism,'' \emph{IEEE Wireless Communications Letters}, vol.~10, pp. 844--848, 2021.

\bibitem{kasgari2018stochastic}
A.~T.~Z. Kasgari and W.~Saad, ``Stochastic optimization and control framework for 5g network slicing with effective isolation,'' in \emph{2018 52nd Annual Conference on Information Sciences and Systems (CISS)}.\hskip 1em plus 0.5em minus 0.4em\relax IEEE, 2018, pp. 1--6.

\bibitem{samanipour2023stability}
P.~Samanipour and H.~A. Poonawala, ``Stability analysis and controller synthesis using single-hidden-layer relu neural networks,'' \emph{IEEE Transactions on Automatic Control}, 2023.

\bibitem{golmohammadi2024lyapunov}
N.~Golmohammadi, M.~M. Rayguru, and S.~Baidya, ``Lyapunov-optimized 5g-sliced communications for telerobotic applications,'' in \emph{IEEE INFOCOM 2024-IEEE Conference on Computer Communications Workshops (INFOCOM WKSHPS)}, 2024.

\bibitem{10.48550/arxiv.2202.06435}
A.~Filali, Z.~Mlika, S.~Cherkaoui, and A.~Kobbane, ``Dynamic sdn-based radio access network slicing with deep reinforcement learning for urllc and embb services,'' 2022.

\bibitem{10.1109/access.2019.2917751}
J.~Zhang, X.~Xu, K.~Zhang, B.~Zhang, X.~Tao, and P.~Zhang, ``Machine learning based flexible transmission time interval scheduling for embb and urllc coexistence scenario,'' \emph{IEEE Access}, 2019.

\bibitem{mahmud2020performance}
M.~H. Mahmud, M.~M. Hossain, A.~A. Khan, S.~Ahmed, M.~A. Mahmud, and M.~H. Islam, ``Performance analysis of ofdm, w-ofdm and f-ofdm under rayleigh fading channel for 5g wireless communication,'' in \emph{2020 3rd International Conference on Intelligent Sustainable Systems (ICISS)}.\hskip 1em plus 0.5em minus 0.4em\relax IEEE, 2020, pp. 1172--1177.

\bibitem{3gpp2017study}
D.~3GPP, ``Study on new radio access technology physical layer aspects,'' \emph{Technical Report (TR) 38.802, V14. 2.0}, 2017.

\end{thebibliography}

\end{document}